# Navigating the Serious Game Design Landscape: A Comprehensive Reference Document

Julieana Moon[1] and Naimul Khan[2]
[1]*Multimedia Research Laboratory, Toronto Metropolitan University, Toronto, Ontario*
*julieana.moon@gmail.com*; *n77khan@torontomu.ca*

**Abstract**

Within the evolving field of digital intervention, serious games emerge as promising tools for evidence-based interventions. Research indicates that gamified therapy, whether employed independently or in conjunction with online psychoeducation or traditional programs, proves more efficacious in delivering care to patients. As we navigate the intricate realm of serious game design, bridging the gap between therapeutic approaches and creative design proves complex. Professionals in clinical and research roles demonstrate innovative thinking yet face challenges in executing engaging therapeutic serious games due to the lack of specialized design skills and knowledge. Thus, a larger question remains: How might we aid and educate professionals in clinical and research roles the importance of game design to support their innovative therapeutic approaches? This study examines potential solutions aimed at facilitating the integration of gamification design principles into clinical study protocols– a pivotal aspect for aligning therapeutic practices with captivating narratives in the pursuit of innovative interventions. We propose two solutions, a flow chart framework for serious games or a comprehensive reference document encompassing gamification design principles and guidelines for best design practices. Through an examination of literature reviews, it was observed that selected design decisions varied across studies. Thus, we propose that the second solution, a comprehensive reference design guide, is more versatile and adaptable.

## 1. Introduction

Serious game is a term that has been used to describe video games that are designed specifically for training and educational purposes [1]. These serious games incorporate gamification techniques by integrating game-like elements into their inventions [18] [34]. The therapeutic use of commercially available video games in various patient populations has been documented in literature since the early 1980s, as reported by [4] and [5]. It wasn't until Ben Sawyer popularized serious games in 2002 when he launched "Serious Games Initiative", a foundation that attracted game designers, educators, and academics alike [2]. The healthcare industry has become increasingly interested in using games as an effective tool for mental health interventions, skill development, and improving patient care [3] [15] [18].

Despite their interests and the significant advancements of serious games, a substantial gap between existing therapeutic treatments and actionable design practices remains. To our knowledge, there are minimal literature reviews that provide a general guide on which gamification techniques or design practices yield best results in serious game design. Current literature reviews suggest design frameworks and principles, often tailored to specific sectors or clinical issues (e.g., educational sector, autism, anxiety, pain management). The existing resources tend to be specialized, lacking a broad overview that would benefit those new to the field or seeking a holistic understanding of serious game design. Thus, there is a gap in addressing novice professionals seeking a high-level overview of the serious game design landscape. Another challenge is the lack of understanding between game designers, developers, and professionals in clinical and research roles. Limited comprehension of team members' roles and industry vertical, in relation to their own expertise, results in suboptimal collaboration and design of serious games [6] [18]. Consider a scenario where a developer, previously specialized in finance applications, is assigned the role of developing a serious game for paediatric patients with anxiety. In this scenario, the developer may lack awareness of design choices that could inadvertently lead to more harm than good. Similarly, professionals in clinical and research roles may find that ideating a serious game is not the most advisable course of action when addressing best practices or efficacy. In other words, serious games have marked a significant milestone in defining what we believe to be the upcoming era of therapeutic technologies, but more bridge building and understanding between supporting members is necessary before implementing design strategies on health.

To aid in this collaboration and help experts from respective professions, we conducted a comprehensive examination of literature reviews on serious games and the broader game design landscape. Our objective was to address the question: How can we effectively communicate to professionals the significance of game design in supporting innovative therapeutic approaches of serious games?

## 2. Methods and Material

### 2.1 Gamification Design Principles Affinity Mapping

Ten gamification design principles from credible literature reviews were revealed following the principles of affinity mapping developed by Jiro Kawakita [7]. The affinity map technique is a structured data visualization exercise used to organize complex qualitative data to understand common themes, identify patterns, and support the meta-analysis interpretation of multiple studies [8]–[10]. Keywords regarding gamification design principles were used as the search terms, resulting in 1,869 results from the Toronto Metropolitan University (TMU) library database [11]–[18] [34]–[43]. Consideration was also given to landscapes of award-wining game studios and their published articles about game design principles [19]–[22]. Thus, twenty-two sources were examined to determine final gamification design principles:

1. Clear Goals and Objectives
2. Design Around Core Mechanics
3. Maintain Game Flow and Player Experience
4. Balancing the Game
5. Storytelling and Immersion
6. Reward and Points System
7. Regular Feedback and Reflection
8. Visual and Auditory Elements
9. Social Affordances
10. Playtesting

### 2.2 Literature Review and Comparisons of Serious Games

Following the refinement of gamification design principles built on the affinity mapping technique, we investigated literature reviews of serious games. Keywords regarding serious games were used in conjunction with *feasibility study, designing and assessing, the effect of, design features of,* and *therapeutic interventions of*. In this section, twelve published literature review of serious games were selected by flexible criteria's:

1. Address any one specific health-related disease or condition
2. Design and develop a serious game or simulation
3. Provide empirical evidence or analysis illustrating the game's efficacy

This exploration focused on comprehending published literature on serious games and evaluating the alignment with our own gamification design principles. Simultaneously, we aimed to investigate if specific clinical protocols were employed in guiding and standardizing the serious game in addressing a particular clinical issue. The exploration further focused on determining whether this had any impact on the application of gamification design principles as presented in Table 1.

**Table 1.** Serious Games Literature Review Summary

| Author(s) (year) | About | Clinical Issue Addressed | Clinical Protocol Used | Used all Game Design Principle? |
|---|---|---|---|---|
| L. Wijnhoven *et al.* (2015) [23] | Mindlight is a computer-based intervention that uses visual aids and structured sensory information to train emotion regulation skills (e.g., relaxation) | Children with an Autism Spectrum Disorder (ASD) and anxiety symptoms | Cognitive Behavioural Therapy (CBT), Neurofeedback, Attention Bias Modification (ABM), Exposure Training | Yes |
| S. Kahlon *et al.* (2023) [24] | Four-armed randomized trial to determine whether gamified Virtual Reality Exposure Therapy (VRET) is more efficacious | Adolescents with public speaking anxiety (PSA) | Cognitive Behavioural Therapy (CBT), Online Exposure Program | No mention of: - Storytelling and immersion |
| A. Ito *et al.* (2023) [25] | Feasibility study of integrating Virtual Reality (VR) into Cognitive Behavioural Therapy (CBT) | Patients with depression, particularly those who do not respond to pharmacotherapy | Cognitive Behavioural Therapy (CBT) | No mention of: - Balancing the game - Regular feedback and Rewards - Maintain Game Flow and Player Experience - Storyline and Immersion |
| C. Jicol *et al.* (2022) [26] | Studies design ethics and efficacy of immersive simulation of street harassment using Virtual Reality (VR) | Street harassment | Multimodal Presence Scale (MPS), Positive and Negative Affect Schedule (PANAS), Intrinsic Motivation Inventory (IMI) | No mention of: - Clear Goals and Objectives - Balancing the Game - Regular Feedback and Rewards |
| G. Pramana *et al.* (2018) [27] | Using mobile health gamification to facilitate cognitive behavioural therapy practices | Children with anxiety disorders | SmartCAT (Smartphone-enhanced Child Anxiety Treatment), Cognitive Behavioural Therapy (CBT) | Yes |
| V. Tashjian *et al.* (2017) [28] | Pain RelieVR is a Virtual Reality (VR) immersive distraction experience | Hospitalized patients with pain, specifically bedbound or limited mobility | Null | Yes |

| N. Davis et al. (2018) [29] | A proof-of-concept study assessing outcomes for a digital treatment game Neuroracer | Children with Anxiety Deficit/Hyperactivity Disorder (ADHD) | Project: EVO (EVO) | Yes |
|---|---|---|---|---|
| E. Ona et al. (2018) [30] | Effectiveness of serious games and the Leap Motion Controller sensors as a tool to support rehabilitation therapies for upper limbs | Patients with Parkison's Disease | Traditional physical therapy exercises: palmar prehension, fingers' flexion, and extension or hand pronation-supination | No mention of: - Storytelling and immersion |
| A. Elnaggar et al. (2016) [31] | Hand rehabilitation using serious games methodology with user-centered design approach | Patients that need hand rehabilitation | Null | Yes |
| P. Kato et al. (2008) [32] | Re-Mission is a PC game used to educate and improve treatment adherence | Children with cancer | Engineered to address behavioural issues identified in literature reviews | Yes |
| P. Siriaraya et al. (2021) [15] | Zen Garden App is a mobile app based on COMET therapy where users plant, grow, and collect resources | Patients with low self-esteem | Competitive Memory Training (COMET) | Yes |
| P. Siriaraya et al. (2021) [15] | The Addiction Beater is music rhythm PC game designed around CBM training modules | Patients with alcohol addiction | CBM training modules (cue-specific response inhibition training and approach bias training) | Yes |

## 2.3 Literature Review of Game Design and Gamification Techniques

Transitioning into the examination of game design and gamification techniques, this section aims to uncover insights and trends that inform our understanding of effective design practices when developing serious games.

Literature reviews suggest different approaches when it comes to design decisions. For instance, A. Schwarz et al.'s work [33] investigates design features associated with engagement in mobile health (mHealth) physical activity (PA) interventions among children, where C. Tziraki et al.'s work [12] investigates design features in patients with dementia (PwD) over 65 years old. A. Shwarz et al.'s study determined that children had positive associations with engagement for features like a clear interface, rewards, multiplayer game mode, social interaction, self-monitoring, and a variety of customization options, including self-set goals, feedback, and progress [33]. Where other features such as sounds, competition, instructions, notifications, virtual maps, or self-monitoring were negatively associated with engagement, emphasizing the need for careful consideration [33]. C. Tziraki et al.'s study, focused on PwD, identified key themes from its serious game pilot study. Firstly, PwD showed a preference for tablets over laptops due to ease of manipulation; they can hold and adjust the device with minimal difficulties [12]. Secondly, auditory cues and instructions were found to enhance performance for PwD, with the delivery modality of instructions being considered crucial considering attention span limitations, potential sensory degradation, and language nuances [12]. As a result, instructions were presented in both written and vocal formats for each game screen [12]. Comparing the two studies, while children in the first study exhibited negative associations with sounds and instructions, geriatric patients using tablets in the second study preferred these features. Significant variations exist in cognitive and emotional development, leading to distinct preferences in the types of games that attract young children compared to those favored by geriatric individuals [11]. The discrepancy in design preferences emphasizes the importance of considering age groups, study designs, and content variations when determining design features [33].

An additional study conducted by [34] offers a comprehensive exploration of game motivators and game principles related to educational serious games. Within this study, 54 educational game design principles in 13 classes and 56 game motivators in 14 classes were presented [34]. This nuanced study builds upon an extensive foundation laid by previous literature reviews on gamification techniques within the educational domain [34].

K Kolln [46] explores a flowchart crafted to facilitate the selection of optimal gamification techniques tailored to a specific use case. A readily accessible process tool that is user-friendly, informative, and suitable for individuals with limited expertise or experience in the field. In her paper "Maybe We Don't Need a New Gamification Framework After All", she elaborates on the idea of an automated process for any use case. Although in the research phase, the aim is to develop a process for finding the most suitable gamification framework (GF) based on factors such as project type, context, budget, or developer skills [46]. It serves as a valuable resource for crafting gamification strategies within a contextually appropriate framework yet comes with some negative connotations. For one, creating a fixed framework may prove challenging and would need extensive research and literature review to support the concrete pathway of a successful serious game. Second, the selection criteria when matching identified frameworks need to be clearly defined when calculating the best solution [46]. Lastly, as improved gamification frameworks continue to emerge, the adaptability and restructuring of the flowchart requires continual refinement to remain current and effective. These factors require meticulous effort and can cause the flowcharts to be unusable.

## 3. Proposed Guidelines and Best Practices

### 3.1 Introduction to Proposed Guidelines and Best Practices

Developers and researchers who seek to create serious games need to consider that design decisions are not universally suitable for all users. A solution that remains flexible, allows for efficient collaboration, and workflow optimization is a comprehensive reference document of gamification design principles and guidelines for best design practices. We consider this solution an optimal choice for effectively conveying the importance of game design because of its adaptability and versatility. The assumption of gamification techniques tied to a specific use case is avoided, affording professionals the freedom to utilize the practical tool as a foundation resource rather than a structural framework.

This document serves as a centralized resource for serious game creation, offering guidance with established design principles and best practices. It streamlines decision-making during the design phase, ensuring a cohesive and standardized approach to serious game development. Derived from literature reviews, research, and practical experience, this reference document stands as an authoritative tool for crafting impactful and engaging serious games.

### 3.2 Gamification Design Principles

Gamification techniques and game design principles are distinct yet interconnected concepts in the realm of game development. Game design principles encompass the fundamental guidelines and elements employed in crafting games, influencing aspects like mechanics, dynamics, and aesthetics. These principles serve as the foundational framework for creating engaging and well-structured gaming experiences [19]–[21]. On the other hand, gamification techniques refer to the strategic incorporation of game-like elements, such as points, badges, rewards, and challenges, into non-game contexts to enhance user engagement and participation [34] [38] [44]. While game design principles focus on the holistic design of games, gamification techniques are specifically applied to non-game scenarios to leverage the motivational and interactive qualities commonly found in games [44]. In essence, game design principles guide the overall construction of games, while gamification techniques extend these principles into various non-game environments, leveraging game-like elements to drive user engagement and achieve specific objectives [38].

Twenty-two literature reviews were evaluated to decide on final gamification design principles [11]–[22] [34]–[43]. The primary focus of these design principles is not directed towards the assessment of the effectiveness, validity, functionality, or success of a serious game. Instead, it centers on providing an overall structure of gamification design principles to explain the learning benefits of and inform the design of serious games.

### 3.2.1 Clear Goals and Objectives

The objective of the game should be clear and concise [22] [34]. It provides a focused direction for the user, enabling coherent and purposeful execution [14] [19] [40] [43]. The serious game may encompass either a primary goal or several small goals. A primary goal is the completion of the entire game, while the small goals serve to motivate the player to progress and achieve the primary goal [19]. Furthermore, the game goal content should be meaningful to the player [20] and should convince the player that there is a deeper meaning in playing beyond entertainment [34]. Goals should be achievable, so the player has a sense of accomplishment.

### 3.2.2 Design Around Core Mechanics

The game should have few core mechanics for the player to immediately understand how to play the game. Nearly every game, whether classified as a serious game or not, revolves around one or few gameplay mechanics [19] [36]. Take Tetris for example, where the core mechanic involves rotating shapes to fit together. Another example is the serious game Re-Mission, their core mechanic is to shoot cancer cells and combat side effects [32]. Defining the core action that players will consistently engage in is crucial to keep players involved [11] [20]. Game controls should be simple to use to lower the learning curve of a game and allow the player to focus on the game's content [19] [21] [34]. One way to achieve this is to use controls that are familiar to the target audience, such as a spacebar for jumping or limiting interaction to one or two buttons.

Game controls should also immediately yield results after input (e.g., the arm of an avatar should mirror the player's arm) [34]. Laggy feedback on input controls may cause lower usability and player disengagement [34]. Another way to design around core mechanics is to use grounded clinical protocols, utilizing real scientific data, to decide on the learning content and activities of the game (e.g., cognitive behavioural therapy programs) [37].

### 3.2.3 Maintain Game Flow and Player Experience

The serious game should maintain the flow of the game and increase player experience. One way to do this is to allow for profile and ownership. Expressive choices refer to decisions made by the player that have minimal impact on learning but can significantly enhance player motivation [42]. These choices might include allowing the player to freely choose an avatar or name their avatar. Granting players control over these aspects can foster the development of empathy and a sense of ownership [42]. Furthermore, offering access to gameplay data (skills learned, goals met, badges awarded, status) through the player's profile is another effective method for promoting ownership and facilitating gameplay adaptation [16] [34].

To maintain game flow, the game's user interface should be simple and intuitive to use. This reduces the cognitive load of the player and is particularly important when target users have impaired cognitive abilities [19] [34]. Another way is to make links to the game's activity and the real world, this increases the players relevance and relatedness to the game [13] [34].

### 3.2.4   Balancing the Game

A game is considered balanced when it presents players with an equal opportunity to succeed [21], a serious game that is easy to learn but difficult to master [20] [22]. It should present challenges aligned with the player's abilities, while avoiding overly complex tasks [19]–[21] [34] [39]. Achieving a balanced game is crucial, as an overly difficult game may demotivate players, while an excessively easy one risks losing the players interest [15] [17] [19] [21] [39]. The progression of levels should begin with instructive tutorials and beginner-friendly challenges, gradually advancing towards more difficult levels [22] [36] [42]. The adjustment of difficulty levels can occur through manual selection (players select their own difficulty level), a preprogrammed system (the difficulty gradually increases as the players progress), or a dynamic approach (players action determine appropriate difficulty level) [34]. Providing adequate time to solve the challenges is also important for enhancing the player's learnability and fostering moments of reflection [34].

### 3.2.5   Storytelling and Immersion

Storytelling is a significant tool for social interaction and information sharing, it enhances player involvement and makes playing more memorable [11] [16] [17] [21] [26] [34] [36]. Narrative transportation, characterized by deep engagement in a story, involves a sense of detachment from reality and a journey into the imaginative realm of the narrative [38]. According to transportation theory, this experience can impact real-world beliefs and actions through three key components: mental imagery, emotional responses, and a diminished connection to real-world information [36] [38]. Stories take place at a specific timeframe, location, and involve characters with external or internal conflicts [11] [39]. To ensure the stories appropriateness, the game's target audience should be clear (e.g., older adult target audiences may not be intrigued by a story about superheroes) [34].

For further immersion, allowing the player to assume the role of a character in the story enhances immersion and expression [17] [42] [43]. The roles can be assigned (the game automatically assigns the role), selected (the player selects a role among pre-created roles), or created (the player creates a role from scratch) [34]. This supports the customization and control motivators [13] [34].

### 3.2.6   Reward and Points System

Rewarding and acknowledging a player's gameplay serves as a motivational factor, encouraging them to engage in the game further [11] [16] [19] [20]–[22] [36]. Psychological studies have shown that rewards serve as a more effective tool for reinforcement than punishment [17]. If a player needs encouragement to perform an action, it is better to use a reward than a punishment (e.g., rewarding a player with powerups after losing all their lives) [17]. Rewards can be in the form of points, achievements, or badges. Points are typically displayed on leaderboards or used to calculate badges/achievements [35].

It's important to note that reward-based gamification presents certain challenges. When aiming for long-term change in serious games, research indicates that relying on rewards can be counterproductive [38] [39]. While this approach is highly effective in achieving short-term change (generates immediate spike in engagement), the discontinuation of rewards can result in the behaviour discontinuing with it [39]. Moreover, reward-based gamification tends to overlook intrinsic motivation, and its excessive use may lead to spill-over effects (a player expects a reward for every task) [38]–[40]. Designers must be mindful that players in such reward cycles anticipate increasing rewards with improved performance, creating a potentially never-ending cycle [39]. If a serious game incorporates extrinsic motivators such as rewards, their purpose should be primarily to offer feedback on player performance, thereby fostering sustained engagement [40]. These rewards should then be replaced with more meaningful intrinsic motivators, such as narrative storytelling, the autonomy to choose paths, engaging activities, and opportunities for reflection [39].

*3.2.7  Regular Feedback and Reflection*

Players experience a sense of competence when provided with distinct feedback for their actions. The feedback should be clear and positive, fostering constructive and educational remarks aimed at improvement [20] [34] [37] [42]. Feedback can be presented through text, visual cues, or auditory cues (e.g., sound effect when a player completes a task) [19] [21] [42].

The concept of reflection involves providing players with opportunities to think about their game-based experiences, a powerful tool in helping a game have meaning well after the game is over [39]. This reflective process allows players to analyze their actions and consider how these choices relate to their own lives. The presentation of reflections can take various forms, ranging from a timeline snapshot of the player's activities to a game character prompting the player to recount their experiences [39] [41]. Alternatively, breaking the fourth wall and having a representative directly engage with the player is another effective method of facilitating this reflective practice [39] [41].

*3.2.8  Visual and Auditory Elements*

Visual and auditory elements in serious games are used to enhance immersive gameplay experience [16] [17] [19] [34] [36] [42]. Visual elements, such as vibrant graphics, intuitive interfaces, and compelling animations, not only make the game aesthetically pleasing but also aid in conveying information, guiding players, and creating an immersive environment [11] [19]. For instance, incorporating visual cues like progress bars, icons, or way-finding arrows can provide players with a clear understanding of their progress or upcoming challenges [36]. On the other hand, auditory elements, including background music, sound effects, and voiceovers, contribute to the emotional engagement of players and add depth to the gaming experience [19]. Well-designed audio can create a more immersive atmosphere, evoke specific emotions, and enhance the overall storytelling within the serious game (e.g., sound cues to reinforce positive actions or accomplishments) [42].

It is also worth mentioning that combining visual and auditory cues can significantly benefit players with cognitive deficiencies, facilitating a better understanding of the game. For instance, individuals with dementia often experience degraded information processing across cognitive, visual, and auditory senses, resulting in reduced cognitive performance [12]. To address these challenges, exploring the integration of multisensory elements is recommended, as it can effectively assist in mitigating physiological changes associated with cognitive deficiencies [12] [16] [37].

*3.2.9  Social Affordances*

Hamari *et al.* classifies guilds, teams, social network, social status, opponents, and direct communication as social affordances serving as channels for interaction and expression [35]. To contribute to a dynamic and engaging social atmosphere, consider incorporating features such as in-game chats, discussion boards, leaderboards, and multiplayer activities [14] [16] [34] [38] [39]. Allowing players to express their status and achievements can significantly influence their in-game actions, acting as powerful motivators for sustained engagement. Furthermore, these social affordances present valuable opportunities for collaboration, enabling players to form groups or engage in activities with friends, fostering a sense of teamwork and shared accomplishments [34] [37] [39].

Creating opportunities for social affordances requires careful consideration of the optimal timing to introduce other players. A player should feel confident within a game environment and its game mechanics before introducing additional players, as premature engagement may make them feel uncomfortable [39]. Forcing a player into a social engagement too quickly can demotivate the player to continue to play [39].

*3.2.10  Playtesting*

Playtesting the serious game with both users and experts can elicit design feedback that reveals the effectiveness of the serious game and the gamification techniques used [13] [19]. Emphasizing design decisions and assessing their impact through a comparative user study can also provide invaluable insights into the efficacy of different approaches, guiding the refinement of future iterations [21] [26].

### 3.3  Guidelines for Best Design Practices

Three significant literature reviews were examined, each providing extensive insights into best design practices. The initial study explores a framework that incorporates therapeutic content into serious games, serving as a valuable resource for novice serious game developers. The second paper highlights the importance of addressing psychological needs when designing technology, emphasizing the human-centric approach. Lastly, the third paper investigates a post-serious game development framework, focusing on ensuring rationale, functionality, validity, and data safety—an essential guide for maintaining the integrity and effectiveness of serious games beyond their initial creation.

While these studies offer valuable frameworks and considerations, the nature of design is inherently dynamic. Professionals are strongly encouraged to adapt and tailor the insights provided by these literature reviews to meet the unique requirements of serious game design. These reviews should be considered as foundational resources, offering valuable perspectives that can be thoughtfully integrated into the design process.

*3.3.1  Framework for Integrating Therapeutic Content into Serious Games*

P. Siriaraya *et al.* [15] introduces a framework to integrate game design into therapeutic content. It proposes a design process framework based on the concept of "game world" versus "therapy world" experiences. It identifies four components of a "therapy world" that can be gamified into a "game world" experience to increase user engagement: performance space, rules, content, and structure (see Fig. 1).

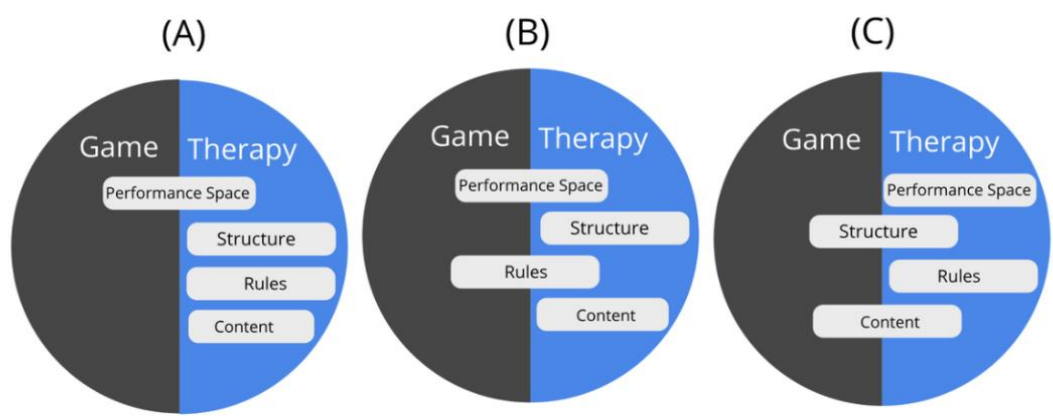

**Figure 1.** Three examples of strategies to integrate game worlds from existing therapeutic activities: (A) Creating a reward system based on the results. (B) Restructuring the rules and performance space. (C) Restructuring the content and structure.

To facilitate the design of gamified therapeutic interventions, the authors propose a Dual-Loop Design model. This model emphasizes the similarities between the core interaction processes in therapeutic activities and game design's core-game loop. The model suggests that a core-game loop can be designed around the therapeutic activity loop, thereby enhancing the user experience of therapy components (See Fig. 2).

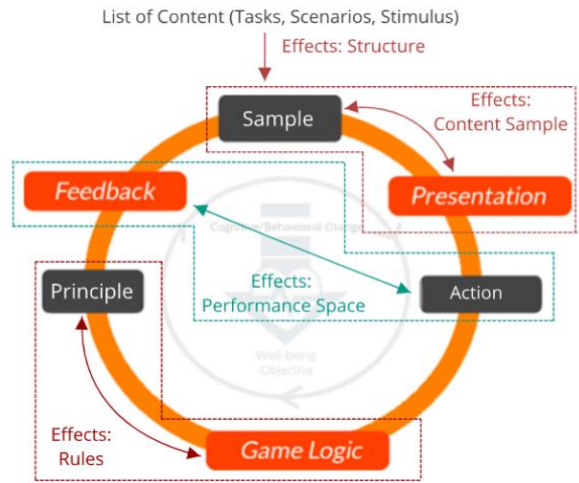

**Figure 2.** Effect of the core-game loop on the therapy game world.

This framework is one of the first papers to combine interdisciplinary health/game research to help designers integrate game design elements with therapeutic content in mind. A key challenge in this integration is balancing the therapeutic and game design components, ensuring that adding gamification techniques enhances engagement without undermining the therapeutic value. The paper argues that merely adding game elements without considering the target audience risks alienating players or undermining the effectiveness of the therapeutic structures. This method was specifically developed to support the design process of serious games, with the key processes used in a therapeutic activity forming the foundation of the concepts and models proposed in a serious game.

*3.3.2    Motivation Engagement and Thriving in User Experience (METUX)*

This paper presents a comprehensive model, Motivation Engagement and Thriving in User Experience (METUX), which integrates psychological research into human-computer interaction (HCI) design [45]. The model emphasizes the importance of satisfying basic psychological needs to foster motivation, engagement, and user wellbeing. These three basic needs are autonomy (feeling agency and alignment with goals and values), competence (feeling able and effective), and relatedness (feeling connected to others, a sense of belonging) [45]. The paper highlights that user engagement in technology is heavily influenced by how technology design aligns with these psychological needs. In other words, if you increase autonomy then engagement will improve, if you increase competence then motivation will increase, and if you increase relatedness then wellbeing will be enhanced [45].

METUX identifies six spheres of influence where technology can impact wellbeing: adoption, interface, task, behavior, life, and society [45]. The model suggests that for technology to be beneficial, it should support psychological needs across these six spheres. For instance, in the adoption phase, user motivation to adopt technology should be autonomous and aligned with their values. In the interface sphere, direct interaction with technology should satisfy psychological needs, enhancing usability and engagement. The task sphere focuses on the fulfillment of psychological needs through specific technology-enabled tasks. The behavior sphere extends to the broader behaviors that technology aims to support or enhance, like exercise or meditation. The life sphere considers the overall impact of technology on an individual's life, including its potential to improve or harm overall wellbeing. Finally, the society sphere looks at the broader societal impacts of technology, such as economic and environmental effects (See Fig. 3).

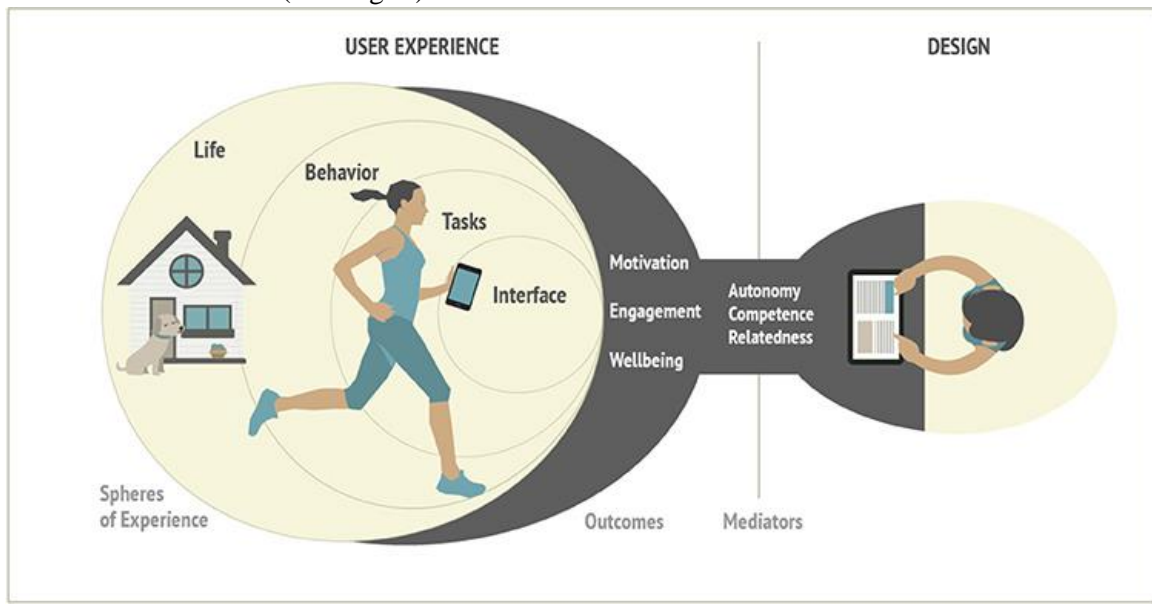

**Figure 3.** METUX model diagram - The basic psychological needs of autonomy, competence and relatedness mediate positive user experience outcomes such as engagement, motivation and thriving.

*3.3.3 Systematically Assess Serious Games*

M. Graafland *et al.* [18] introduces a consensus-based framework for assessing medical serious games, providing sixty-two items across five main themes: rationale, functionality, validity, and data safety. It was developed by the Dutch Society for Simulation in Healthcare (DSSH) and based on the reporting standards for non-game mobile health apps (mHealth) [18]. The framework was developed to address the unique aspects of serious games and how they differ from standard mHealth apps in terms of content and interaction goals. Overall, this framework provides a structured approach for evaluating serious games in healthcare, supporting all professionals involved in assessing the relevance, validity, and safety of the game. It helps inform decisions when applying a serious game for healthcare purposes.

## 4. Conclusions

We began this work operating under the assumption that professionals in clinician and research roles have not thoroughly investigated the gamification design landscape and best design practices. The effectiveness of serious games to change health behaviors and improve patient care suggests that the building blocks of serious games should be considered when designing interventions in healthcare [32].

This literature review revealed that there is much more to be explored from gamification techniques in serious games and has the potential to be combined with earlier research to create new research paths. The intent of this research was not to offer a definitive list of gamification techniques or to advocate implementing these into serious games. Instead, this investigation was constrained to examining how different gamification techniques from literature reviews might help inform professionals about the creation of serious games. Rather than picking and choosing gamification techniques out of a bag, we suggest researchers and relevant professionals collaborate and use the fundamental elements of gamification design principles as a guide. We expect user testing and controlled studies to further help in the developmental phase of serious games, bringing light to the efficacy of chosen gamification techniques.

# Acknowledgments

The authors would like to acknowledge funding from the Government of Canada through the New Frontiers in Research Funds program.

# Conflicts of interest

The Authors have no conflicts to declare.